\UseRawInputEncoding
\documentclass[pdflatex,sn-mathphys-num]{sn-jnl}

\usepackage{graphicx}%
\usepackage{multirow}%
\usepackage{amsmath,amssymb,amsfonts}%
\usepackage{amsthm}%

\usepackage{mathrsfs}%
\usepackage[title]{appendix}%
\usepackage{xcolor}%
\usepackage{textcomp}%
\usepackage{manyfoot}%
\usepackage{booktabs}%
\usepackage{algorithm}%
\usepackage{algorithmicx}%
\usepackage{algpseudocode}%
\usepackage{listings}%
\usepackage{float}

\definecolor{mycolor}{RGB}{250,250,0}
\usepackage{soul}

\theoremstyle{thmstyleone}%
\theoremstyle{thmstyletwo}%

\theoremstyle{thmstylethree}%

\begin{document}

\title[Article Title]{Quadrature-Dependent Lattice Dynamics of  Dissipative Microcombs
}


\author*[1]{\fnm{Eran} \sur{Lustig}}\email{elustig@stanford.edu}
\equalcont{These authors contributed equally to this work.}
\author[1]{\fnm{Melissa A.} \sur{Guidry}}
\equalcont{These authors contributed equally to this work.}
\author[1]{\fnm{Daniil M.} \sur{Lukin}}
\author[1]{\fnm{Shanhui} \sur{Fan}}
\author[1]{\fnm{Jelena} \sur{Vu\v{c}kovi\'{c}}}
\affil[1]{\orgdiv{Edward L. Ginzton Laboratory}, \orgname{Stanford University}, \city{Stanford}, \postcode{94305}, \state{CA}, \country{USA}}

\abstract{The study of coupled networks with parametric amplification of the vacuum fluctuations has garnered increasing interest due to its intricate physics and potential applications \cite{mcdonald_phase-dependent_2018,wang_non-hermitian_2019,wan_quantum-squeezing-induced_2023,flynn_deconstructing_2020,wanjura_quadrature_2023,javid_chip-scale_2023}. In these systems, parametric interactions lead to beam-splitter coupling and two-mode squeezing, creating quadrature-dependent dynamics. These systems can be modeled as bosonic networks, arrays or lattices, exhibiting exotic effects such as unidirectional amplification and non-Hermitian chiral transport which influence multimode squeezing \cite{del_pino_non-hermitian_2022,slim_optomechanical_2024,wanjura_topological_2020}. However, exploring and controlling these network dynamics experimentally in all-optical systems remains challenging. Recent advancements in integrated nonlinear micro-resonators, known as Kerr microcombs, enable the generation and control of broadband high-repetition pulses on microchips \cite{delhaye_optical_2007,herr_temporal_2014,cole_soliton_2017}. Kerr microcombs exhibit intriguing nonlinear dynamics, where coherent photons occupy discrete spectral lines, leading to multimode squeezed vacuum states \cite{guidry_quantum_2022,guidry_multimode_2023,gouzien_hidden_2023}.
In this work, we explore the lattice dynamics of vacuum fluctuations driven by dissipative Kerr microcombs. We design a photonic chip where a spontaneously emergent pair of pulses creates extended multimode states of parametrically amplified vacuum fluctuations. These states exhibit oscillatory dynamics, with implications on squeezing and secondary comb formation. By employing integrated micro-heaters, we tune the vacuum fluctuations to eliminate the oscillations, establishing a fundamental connection between non-Hermitian lattice symmetries and Kerr combs, paving the way for exotic quadrature-dependent optical networks with broad implications for quantum and classical photonic technologies.}

\maketitle

\section{Introduction}\label{sec1}

In recent years, there has been a growing interest in exploring the dynamics of multimode driven vacuum fluctuations \cite{mcdonald_phase-dependent_2018,wanjura_topological_2020}. Traditionally, quantum vacuum fluctuations are parametrically amplified and subsequently undergo beam-splitting operations, leading to rich and potentially useful multimode squeezed states \cite{de_valcarcel_multimode_2006,roslund_wavelength-multiplexed_2014,chen2014experimental,cai_multimode_2017,madsen_quantum_2022}. However, when parametric amplification and transport between different modes occur simultaneously, they induce dynamics of coupled networks or lattices, presenting a plethora of new effects from the perspective of squeezing and deterministic entanglement generation \cite{wan_quantum-squeezing-induced_2023,flynn_deconstructing_2020,mcdonald_phase-dependent_2018,poli_selective_2015,luo_quantum_2022,uddin_quantum_2024}. 
These networks can be studied within the framework of the Bogoliubov modes and bosonic linear quadratic Hamiltonians. The eigenstates of the system, referred here as supermodes, are represented in the quadrature basis \cite{fabre_modes_2020}, which is also the natural basis for continuous-variable (CV) quantum states \cite{RevModPhys.77.513}. The quadratic bosonic Hamiltonians here exhibit non-trivial non-Hermiticity which is not from dissipation, as the underlying Hamiltonian is Hermitian. Nevertheless, the parametric processes give rise to non-Hermitian quadrature-dependent dynamics \cite{wang_non-hermitian_2019,flynn_deconstructing_2020,roy_spectral_2021}. Such processes were recently explored in optomechanics within a synthetic frequency dimension \cite{del_pino_non-hermitian_2022,slim_optomechanical_2024}, exemplifying how symmetries and transport play a crucial role in the dynamics of the squeezed vacuum.

In the optical domain, multimode amplified quantum vacuum appears in both second-order and third-order nonlinear systems \cite{chembo_quantum_2016,yang_squeezed_2021,zhao_near-degenerate_2020,jahanbozorgi_generation_2023}. In second-order nonlinear systems, photon transport between modes does not occur naturally and requires additional modulation to achieve lattice transport \cite{javid_chip-scale_2023}. In contrast, third-order nonlinear processes present a different scenario. Remarkably, Kerr microcombs \cite{delhaye_optical_2007}—including dissipative Kerr solitons \cite{herr_temporal_2014} and soliton crystals \cite{cole_soliton_2017}—generated by third-order nonlinear resonators naturally produce both parametric amplification \cite{reimer_generation_2016, kues_-chip_2017} and photon transport between resonator modes through the Bragg scattering process \cite{chembo_quantum_2016,zhao_near-degenerate_2020,moille_synthetic_2022,englebert_bloch_2023,guidry_quantum_2022}. This makes Kerr combs a unique system for studying multimode quantum phenomena \cite{guidry_quantum_2022, bensemhoun2024multipartite} and multimode squeezing \cite{guidry_multimode_2023,gouzien_hidden_2023}.

Consequently, Kerr combs can potentially induce different network geometries and lattice dynamics of multimode amplified vacuum fluctuations. This property of Kerr combs not only presents new methods for generating and controlling multimode quantum light but also enhances our understanding of classical nonlinear state evolution through the interplay between driven vacuum fluctuation dynamics and Kerr comb evolution in the resonator.

In this study, we investigate the lattice dynamics of optical vacuum fluctuations driven by dissipative Kerr micro-combs. We designed a microchip in which a pair of emerging pulses parametrically amplify and induce transport of the vacuum fluctuations, which subsequently exhibit lattice dynamics. We observe that the vacuum fluctuations generally oscillate, and that these oscillations are associated with a mismatch between the micro-comb repetition rate and the underlying dispersion. By utilizing integrated thermo-optic tunability, we are able to completely eliminate the oscillations and reach a stable, non-oscillatory (or stationary) regime of the amplified vacuum fluctuations. To explain this phenomenon, we draw parallels with classical parity-time (PT) symmetric lattices \cite{ruter_observation_2010,regensburger_paritytime_2012,zeuner_observation_2015,weimann_topologically_2017}. Furthermore, we demonstrate how the oscillatory nature of the amplified vacuum fluctuations affects threshold processes, leading to the onset of (or lack of) RF beat notes and subsequent chaos in our micro-resonator. This advancement opens the door for modeling and utilizing bosonic network dynamics in an all-optical integrated platform with implication on multimode squeezing and entanglement between many frequency modes.

\section{Results}\label{sec2}

\begin{figure}[ht]
\centering
\includegraphics[width=1\textwidth]{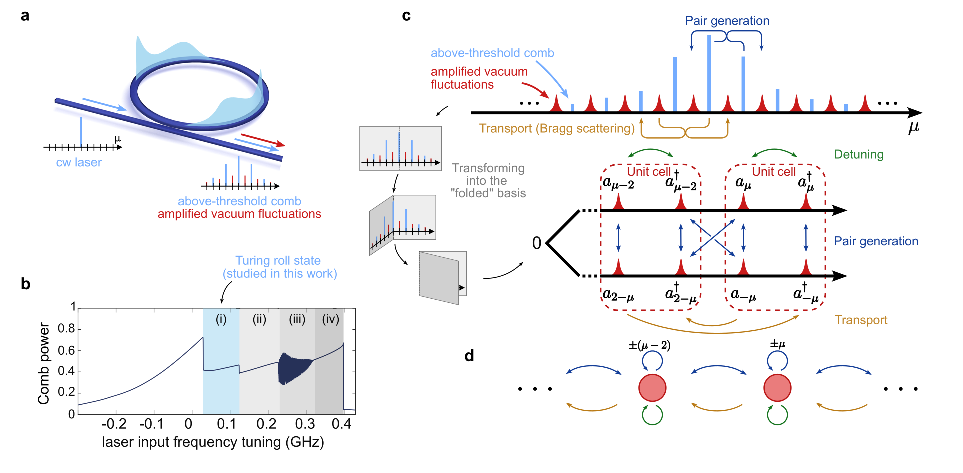}
\caption{ \textbf{Scheme for exploring the lattice dynamics of vacuum fluctuations driven by dissipative Kerr microcombs}. \textbf{a.} We inject a CW laser into a high-Q microring, which generate dissipative Kerr combs, and we measure the transmitted light. \textbf{b.} The nonlinearity and dispersion of our devices lead to several nonlinear transitions, as shown in the simulated comb power plot vs input laser frequency detuning, a solution of the Lugiato-Lefever equation for representative system parameters. Our main focus is the state (i) after threshold, in which two Turing roll pulses (forming a 2-FSR primary comb) spontaneously emerge, and the threshold after it leading to (ii) a secondary comb. Following thresholds lead to (iii) transition oscillations, and finally (iv) solitons.
\textbf{c.} The spectrum of the two pulses shows that between each pair of neighboring comb lines (even modes), there is a resonance with amplified vacuum fluctuations (odd modes). The pulses induce transport (Bragg scattering), pair-generation (parametric amplification), and detuning (cross phase modulation) of the vacuum fluctuations. The lower panel shows how folding the frequency axis reveals lattice dynamics defined by a unit-cell of the vacuum fluctuations. The relatively narrow bandwidth of our pulses means that transport is governed by nearest-neighbor terms, which result in the lattice dynamics of the vacuum fluctuations.  \textbf{d.} A simplified scheme of the lattice geometry of the lower panel of (c), each red circle represents a single unit-cell composed of two cavity modes}\label{fig1}
\end{figure}

To study the lattice dynamics of vacuum fluctuations driven by dissipative Kerr combs, we inject single-frequency (continuous wave, CW) light into a high-Q micro-resonator made of thin-film silicon carbide on insulator (more on the device in Methods~\ref{Methods1}) and spontaneously generate dissipative Kerr microcombs (Fig.\ref{fig1}(a)). By red-tuning the input CW laser beam toward a resonance, a sequence of nonlinear transitions occurs, leading to the formation of different microcombs. Figure~\ref{fig1}(b) shows the Lugiato–Lefever equations (LLE) simulation \cite{chembo_spatiotemporal_2013} of the power inside the resonator and corresponds to our device (see Methods~\ref{Methods3} for more details on the simulation). The discontinuous steps in the transmission are nonlinear state transitions which correspond to different comb spectra (see Methods~\ref{Methods2} for experimental comb spectra). For now, we focus on the first generated comb - the primary comb which appears after the first nonlinear threshold. Figure~\ref{fig1}(c) illustrates that in our device this microcomb corresponds to the spontaneous formation of two pulses via the symmetry-breaking phenomenon of Turing rolls \cite{coen2001continuous}. Accordingly, the spectrum of our two pulses forms a frequency comb with discrete frequency lines separated by twice the free spectral range (FSR), which is also the pulse repetition rate (blue lines in Fig.\ref{fig1}(c)). In our system the FSR is 150 GHz.

To experimentally achieve the spontaneous formation of combs that include multiple frequency lines within a relatively narrow bandwidth (1 THz), we required strong anomalous dispersion. The dispersion of the material forming the waveguide is normal, and the sufficiently strong anomalous dispersion could not be achieved by geometric dispersion via tuning the waveguide cross section. Instead, we chose a multimode cross section and selected devices which feature an avoided mode crossing of two mode families - transverse electric (TE)/ transverse magnetic (TM). The mode crossing strongly detunes the resonance close to the pump frequency (Methods~\ref{Methods1}), which allows for an optical parametric oscillation (OPO) threshold between cavity modes that are separated in frequency by only 300 GHz, which is twice the FSR. This leads to the emergence of the 2-FSR primary Kerr comb, as illustrated in Fig.~\ref{fig1}(c).

The choice of a 2-FSR primary Kerr comb with a relatively narrow bandwidth, confine the multimode dynamics and simplify the geometry of the driven vacuum fluctuations. The driven vacuum fluctuations that are the focus of our study are in the modes that are not populated with coherent light. To identify these modes we assign numbers $\mu$ to the resonances in our cavity. We define the pumped resonance as $\mu=0$ and thus the 2-FSR microcomb exists on every even-numbered cavity mode which are predominantly populated by coherent light. In contrast, light in the odd-numbered modes is dominated by vacuum fluctuations which are driven (pair generated) by the microcomb (indicated by red areas in Fig.~\ref{fig1}(c)). This occurs as long as the cavity loss in the odd-numbered modes is greater than the parametric gain generated by the microcomb. The light remains in a driven vacuum state until the parametric gain increases above the cavity loss leading to a secondary threshold that generates more coherent comb lines. 
Next, we discuss the connectivity of the driven vacuum fluctuations and their modeling as a network of sites (cavity modes) and more specifically: a lattice. Figure~\ref{fig1}(c,d) present the network model for our below threshold light in the odd-numbered modes, which has a lattice geometry. Transport, which is the result of Bragg scattering, connects primarily odd cavity modes to their neighboring odd cavity modes rather than to the even-numbered modes. This aspect makes our network into a one-dimensional lattice rather than a more general network of cavity modes populated by amplified vacuum states of light. 

The lattice geometry is also related to the pair generation in our system. In Fig.~\ref{fig1}(c) (lower panel), we plot the frequency axis in a ``folded'' form. In the folded form, modes $-\mu$ and $\mu$ are brought close together. The pair generation process is driven primarily by the pump mode $\mu=0$ together with the side-bands in cavity modes $\mu=\pm2$ (see section 1 in the Supplementary Material). From energy conservation, pair generation occurs primarily between modes that are in proximity to one another within the folded geometry. Thus, if a photon is generated (or annihilated) in mode $\mu$ it's pair will likely generate (or annihilate) at modes $-\mu, (-\mu+2)\text{ or } (-\mu-2)$. From similar considerations, Bragg scattering will be governed by transport of photons from mode $\mu$ to a close mode at $\mu=(\mu+2)\text{ or } (\mu-2)$. The simplified lattice model in the folded geometry is plotted in Fig.\ref{fig1}(d). Taking into account all interactions, we can define a periodic unit cell consisting of four bosonic field operators: $a_\mu$, $a^\dagger_\mu$, $a_{-\mu}$, and $a^\dagger_{-\mu}$. The unit cell is the defining feature of the lattice model.  

In what comes next, we solve the dynamics numerically, without relying on the simplified lattice model presented in Fig.\ref{fig1}(c-d). The dynamics of the below-threshold light at the odd-numbered modes, centered around the pump at $\mu = 0$, is governed by the following linearized Hamiltonian \cite{guidry_quantum_2022}.
\begin{equation}
\
\hat H=\sum_{\mu}-\Delta\omega_{\mu}\hat{a}_{\mu}^{\dagger}\hat{a}_{\mu}-\frac{g_{0}}{2}\sum_{\mu,\nu,j,k}\delta\left[\mu+\nu-j-k\right]\left(A_{\mu}A_{\nu}\hat{a}_{j}^{\dagger}\hat{a}_{k}^{\dagger}+2A_{k}^{*}A_{\nu}a_{j}^{\dagger}a_{\mu}+\text{h.c}\right)
\label{eq1}
\end{equation}

where $A_\mu$ are above-threshold amplitudes (the Kerr comb) in even cavity modes $\mu$, $\hat{a}_\mu^\dagger \left(\hat{a}_\mu\right)$ are the creation (annihilation) operators of quantum light in odd cavity mode $\mu$, which are described in the rotating frame of reference of the Kerr comb $U=\exp(i\hat{R}t),\;\hat{R}=\sum_\mu\left({\omega_p+\Delta\Omega \mu }\right)a^\dagger_\mu a_\mu $ where $\Delta\Omega$ is the frequency spacing (repetition rate) of the Kerr microcomb. $\Delta \omega_\mu$ is the frequency detuning of the cavity-mode $\mu$ from the Kerr comb rotating frame of reference, $g_0$ is the nonlinear coefficient, the Kronecker $\delta$ reflects photon azimuthal wave number conservation in the system, and h.c stands for Hermitian conjugate. 

The Hamiltonian in Eq.~\ref{eq1} can be numerically solved for a given microcomb amplitudes $A_\mu$ by obtaining the covariance matrix associated with a discrete set of supermodes \cite{flynn_deconstructing_2020}, which in our case extends across multiple odd numbered resonances. Each supermode is characterized by a complex eigenvalue, where the real part is the parametric gain, i.e., the squeezing and anti-squeezing of the light, and the imaginary part is the oscillations of the supermode. The complex eigenvalues comes from the non-Hermitian dynamical matrix $\mathcal{M}$ that describes the quadrature dependent evolution in time. $\mathcal{M}$ is calculated by writing the Hamiltonian in Eq~\ref{eq1} in the quadrature basis $\vec{R}=(p_{-N},...,p_{N},q_{-N}...,q_{N})^T$ where $p_\mu=\frac{1}{\sqrt{2}}\left(a_\mu^\dagger+a_\mu\right),q_\mu=\frac{i}{\sqrt{2}}\left(a_\mu^\dagger-a_\mu\right)$, and defining $\mathcal{M}$ as the matrix that satisfies the Heisenberg equation: $\frac{\partial\vec{R}}{\partial t}=\mathcal{M}\vec{R}\label{eqM1}$. The quadrature dependent dynamics described by $\mathcal{M}$ carry some analogies to classical non-Hermitian lattice \cite{wang_non-hermitian_2019}, as we will further discuss in this work. 

In principle, Bragg scattering can generate extended lattices and supermodes over arbitrarily large spectral distances. In our system the mechanism that limits the spectral distance and hence the supermode spectral bandwidth is the anomalous dispersion ($D_2/2\pi = 1.2~\text{MHz}$, where $D_2$ is the second-order dispersion) of our cavity (See Methods \ref{Methods6} for the finite model numerical analysis). In this context, it is noteworthy that the cavity modes in our system function as a lattice in the synthetic frequency dimensions  \cite{yuan_photonic_2016,ozawa_synthetic_2016}, where transport is induced by Bragg scattering. The supermodes are essentially Floquet lattice modes at twice the repetition rate frequency \cite{lustig_photonic_2019,dutt_experimental_2019,wang_generating_2021} confined by $D_2$, which acts as an effective confining potential \cite{dutt_creating_2022}.

We begin the experimental investigation of the below-threshold light by measuring the RF spectrum of a single extended supermode on the odd cavity modes. Our measurements are performed close to the OPO threshold, which means that at least one supermode has high gain that almost compensates the cavity loss. Close to the OPO threshold, typically one or two supermodes are dominant and amplified above all others. To enable detection of vacuum fluctuations in the presence of the relatively intense Kerr microcomb in the cavity, we utilize both on-chip and off-chip filtering (see Methods~\ref{Methods3}). Such filtering is made possible due to the large FSR (~150~\text{GHz}) between resonances in our micro-resonator.

\begin{figure}[ht!]
\centering
\includegraphics[width=1\textwidth]{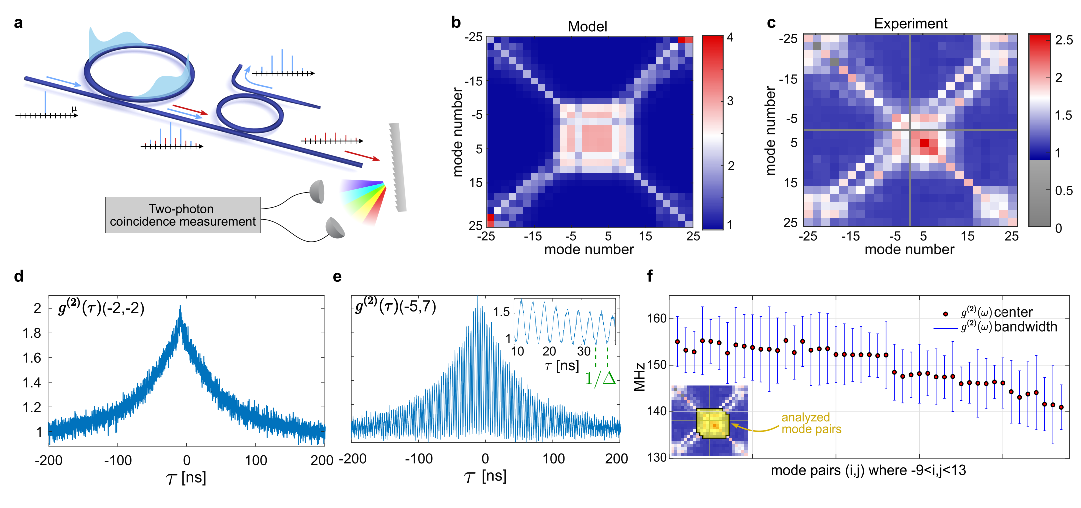}
\caption{ \textbf{Resolving the dominant supermode correlations and oscillations near threshold} \textbf{a.} Schematics of measuring the parametrically amplified vacuum state correlations: the light in the odd cavity modes is filtered on-chip (filter ring) and off-chip (monochromator). Subsequently, measurement of the photon populations below threshold is obtained using SPOSA. \textbf{b.,c.} The two-photon correlation ($g^{(2)}(t=0)$) matrix of the below threshold light in the odd-numbered modes, in simulation and experiment respectively. We observe all-to-all correlations between cavity modes $-9$ and $13$ (central square in the correlation plot) generated by Bragg scattering and pair-generation. \textbf{d.} Auto correlation of mode $-2$ as a function of temporal delay. The measurement is for single pump without a microcomb in the resonator. The large temporal width of the signal indicates the proximity to threshold. \textbf{e.} Cross correlation as a function of temporal delay of the mode pairs $(-5,7)$ when driven by the micro-comb. The carrier frequency $\Delta$ appears in contrast to the case in (d.) where it does not appear. \textbf{f.} The frequency carrier $\Delta$ and the frequency bandwidth (inverse of the temporal broadening) of the auto and cross correlations of the pairs of odd-numbered modes between -9 and 13 marked in yellow in the inset. The uniformity of the bandwidth and oscillations $\Delta$ indicates the dominance of a single supermode with a single complex eigenvalue. The data is sorted according to acquisition time, showing that the main deviations from a constant frequency is due to the hot cavity dispersion gradually changing with time.}\label{fig2}
\end{figure}

We use a custom-built single-photon optical spectrum analyzer (SPOSA) \cite{guidry_quantum_2022}, which provides $>100$~dB of dynamic range with single-photon sensitivity, to simultaneously capture the photon populations above and below threshold (see Methods~\ref{Methods2}). Mapping out the two-photon correlation matrix of the below-threshold state can reveal the inter-modal connectivity in the Hamiltonian and provide indications of the formation of the supermodes. We perform pairwise second-order photon correlation measurements, $g^{(2)}_{ij}(\tau) = \langle \hat{a}_i^\dagger(0) \hat{a}_j^\dagger(\tau) \hat{a}_j(\tau) \hat{a}_i(0) \rangle / {\langle\hat{a}_i^\dagger(0) \hat{a}_i(0) \rangle \langle\hat{a}_j^\dagger(\tau) \hat{a}_j(\tau)\rangle}$, on each pair of below-threshold modes (Fig.~\ref{fig2}(a)). The resulting matrix (both simulation and experiment) is shown in Fig.~\ref{fig2}(b-c), respectively. The most notable features is the presence of a square near the center of the correlation map. This square is related to the presence of a dominant single extended supermode as will be discussed in more details below. The square indicates all-to-all correlation in a finite set of coupled resonances. In our case, the group velocity dispersion limits the lattice dynamics to an area of 12 resonances (odd-numbered modes between -9 and 13), which is the size of the square.

The square shape observed in the cross correlation data in Fig.~\ref{fig2}(c) suggests the presence of extended all-to-all correlations across 12 cavity modes, resulting from photon transport and pair generation (See section 1 in the Supplementry Material). As mentioned, close to the OPO threshold there should be a single dominant supermode. This allows us to identify the RF spectrum of the single extended supermode that reached threshold (which corresponds to the eigenvalue of that supermode). To resolve the RF spectrum, we exploit a notable feature of the multimode vacuum fluctuations generated by microcombs - they oscillate close to threshold. This can be seen by comparing the time-dependent correlations \( g^{(2)}_{i,j}(\tau) \) of squeezed vacuum produced by a CW laser near threshold (Fig. \ref{fig2}(d)), with vacuum fluctuations amplified by multi-frequency light (Fig. \ref{fig2}(e)). The key distinction between Fig. \ref{fig2}(d) and Fig. \ref{fig2}(e) is the presence of an underlying RF beat note in the multimode scenario. Mathematically, these oscillations represent the imaginary part of the eigenvalue of the dominant supermode. The appearance of the oscillations in the time-dependent correlation of Kerr combs close to threshold was theoretically predicted in \cite{guidry_multimode_2023}, and we confirm them experimentally in this work. 

Before examining the origin of the oscillations of the supermode near threshold, we demonstrate that the oscillations are shared by all auto-correlations and cross-correlations among the central 12 resonances, consistent with a single dominant supermode with a complex eigenvalue. Figure \ref{fig2}(f) presents the oscillations $\Delta$ (red data point) and the bandwidth (range in blue) of the auto-correlations and cross-correlations in the odd cavity modes from cavity modes -9 to 13. All of the auto-correlations and cross-correlations exhibit frequency beating in the range of \(\Delta = 140-160\) MHz, with a bandwidth of approximately 10 MHz. This bandwidth is an order of magnitude narrower than the resonance linewidth and more than an order of magnitude broader than the laser linewidth which shows that all modes are below the OPO threshold and are consequently squeezed by the optical drive. 
 
By ordering the measured frequencies in Fig. \ref{fig2}(f) according to acquisition time (see data file 1), spanning several hours, it is evident that the majority of the non-uniformity between the different pairs of modes comes from gradual shifts in time in the hot-cavity dispersion. As mentioned, each supermode's complex eigenvalue consists of a real part, which represents the parametric gain (and correspondingly the bandwidth, as indicated by the blue bars in Fig. \ref{fig2}(f)), and an imaginary part, which represents the oscillations (depicted by the red dots in Fig. \ref{fig2}(f)). Therefore, the consistency of the oscillations and the bandwidth across various correlations demonstrates that light in these cavity modes correspond to the same eigenvalue. If there were two supermodes, their distinct frequencies would create a fast beatnote close to threshold. This effectively resolves the RF spectrum of a single extended supermode of an emerging dissipative microcomb near threshold.

Next we study the evolution of the supermode as a function of proximity to threshold. By tuning closer to resonance, we amplify the oscillating supermode described in Fig.~\ref{fig2} such that the parametric gain of one supermode exceeds the loss resulting in a secondary threshold. By performing multi-frequency homodyne measurements as illustrated in Fig.~\ref{fig3}(a) (see also Methods~\ref{Methods4} for the experimental configuration and section 4 in the Supplementary Material) we are able to measure multimode anti-squeezing of the vacuum fluctuations associated with the transition from Turing rolls to multiple RF beatnotes (Fig.~\ref{fig3}(b)). The secondary threshold leads to the appearance of multiple classical RF beat notes (in addition to the two peaks shown in the left upper panel Fig.~\ref{fig3}(b)). While the onset of beat notes was studied in \cite{herr_universal_2012}, observing the vacuum fluctuations before their appearance sheds additional light on this physical phenomenon. The onset of multiple classical RF beat notes after a secondary threshold may be attributed to multiple incommensurate combs broadening and merging together after threshold \cite{herr_universal_2012}. By tracking this phenomenon in the below-threshold regime, we observe that in our case, onset of oscillations arises from a single oscillating supermode, which arises from the incommensurate nature of vacuum fluctuations below threshold.

The theoretical plot in Fig.~\ref{fig3}(b) illustrates that the oscillating supermode becomes a coherent RF beat note after crossing threshold, which then generates additional beat notes, eventually leading to lines filling the entire spectrum. Measurements in Fig.~\ref{fig3}(c) confirm this by tracking the noise variance of each resonance through the process of threshold crossing. By adjusting the local oscillator to correspond with different cavity modes, we observe a uniform RF beat note identical to that of the below-threshold light in Fig.~\ref{fig2}. At the threshold, the entire supermode increases in intensity and narrows in bandwidth to generates a uniform coherent beat note. In the same manner, the $g^{(2)}(\tau)$ correlations broaden (see Section 5 in the Supplementary Material). Even before reaching the OPO threshold, the explanation given here is simplified, since the supermodes oscillate at frequencies that can vary as the microcomb evolve with the change of proximity to threshold. While these variations in principle can be large, in our case here, they are observed to be small. At the point of the secondary OPO threshold, the experimental results (upper plots in Fig.~\ref{fig3}(c)) display that the below threshold oscillations turn into a classical beat note that generate with the microcomb new beat notes. The process of generating new beat notes is the onset of chaos and we will show how to prevent it by thermally tuning the below-threshold light. 

\begin{figure}[ht]
\centering
\includegraphics[width=1\textwidth]{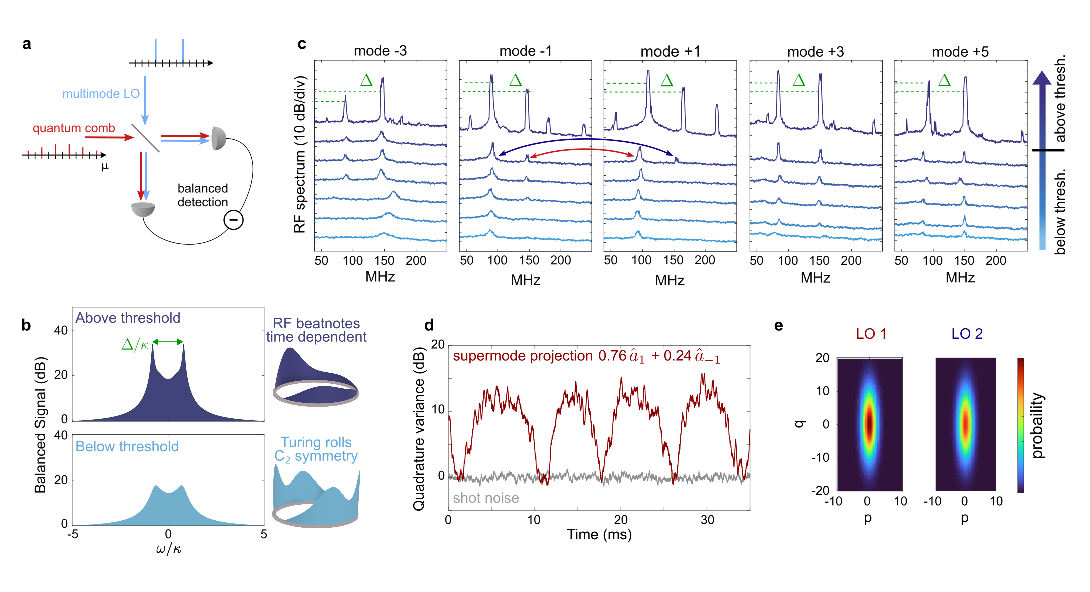}
\caption{{{\textbf{Tracking the supermode dynamics: from oscillating driven vacuum fluctuations to multiple RF beat notes} \textbf{a.} The transition is tracked using homodyne detection with a multi-frequency local oscillator. \textbf{b.} An illustration of the microcomb transition in the RF spectrum (left) and in the azimuthal intensity distribution (right): on the right: the 2-FSR Turing roll pulses evolve into a state characterized by multiple RF beat notes, leading to chaos. on the left: the RF squeezing and anti-squeezing spectrum of the supermode showing the progression from an oscillating extended supermode to a coherent beating state that generates additional beat notes. \textbf{c.} Spectroscopy measurements at the odd cavity modes (local oscillator frequency corresponds with mode $\mu$, gradually approaching the secondary threshold (indicated by color), with classical beat notes appearing above threshold. \textbf{d.} Multi-frequency homodyne detection is performed on cavity modes -1 and 1 simultanously, with two optimal amplitude ratios $\beta$ (blue and red) to maximize the antisqueezing. Each pair of peaks leads to a quadrature variance that reaches shot noise near threshold. \textbf{e.} The reconstructed quadrature probability of the oscillatory supermode, based on pairs of peaks (LO1 in red and LO2 in blue correspond with optimizing the LO for the peaks with the red and blue arrows in (c.))}}}\label{fig3} 
\end{figure}

Before moving to our control of the oscillations, we wish to observe the quadrature variance for different quadratures in the double peak structure that we measure with our noise spectroscopy detection in Fig.~\ref{fig3}(c). Since the supermodes are below threshold the variance of one of the quadratures is at or under shot-noise level. To show this, we perform homodyne measurements using a local oscillator composed of two different frequencies corresponding to resonances: $\mu=-1,1$. This configuration allows for much better detection of the different quadrature variances than a single frequency in the local oscillator (while not requiring an active phase locking between the local oscillator and the microcomb \cite{marino_bichromatic_2007}). Figure \ref{fig3}(d) presents the  quadrature-dependent variance of the oscillating supermode as we temporally sweep the phase of the local oscillator (LO). Modes -1 and 1 each contain four peaks, which are part of the approximately 24 peaks in the entire supermode. To observe the below-threshold quadrature dependent variance, we adjust the power ratio of our two-tone LO for modes -1 and 1 to $LO_1 = \beta \cdot LO_{-1}$, where $\beta$ is a ratio that maximizes the overlap with a projection on the amplified quadrature supermode. Without using at least two-tones in our homodyne detection, and adjusting their ratio, the oscillation variance would not dip to shot-noise levels for any phase. Figure \ref{fig3}(e) shows the reconstructed Wigner quasi-probability distribution associated with the pairs of peaks marked by red and blue arrows in Fig.\ref{fig3}(c). Each pair requires it's own optimized LO: LO1 and LO2 ($\beta_1=0.1,\beta_2=0.16$). The q and p in Fig.\ref{fig3}(e) are the quadratures that correspond to the optimized LO1 and LO2 for each subplot. These pairs are not separated by the native spacing of the microcomb in the even-numbered modes. In performing these measurements, we showed how a below-threshold supermode transition to classical light, and impact the secondary comb formation.

After showing that the oscillations of the supermode affect the below threshold quadrature variance, and the comb evolution, it is important to understand the oscillations origin and control them. In our system, Bragg scattering primarily transports photons between resonances separated by a frequency spacing of 2-FSR, generating the depicted lattice dynamics in our below-threshold light (see Fig. \ref{fig1}(c-d)). It turns out that the below-threshold oscillations we observe can be attributed to the absence of anti-parity time symmetry in the lattice model (see Methods \ref{Methods5}). In scenarios involving squeezing originating from a single pump, or in single-mode squeezing, anti-parity time symmetry is inherent. For this reason oscillations are not commonly observed in squeezing experiments near threshold (Fig. \ref{fig2}(d)). However, in the multi-pump, multi-mode scenario that we explore here, the presence of two different frequency spacings: the repetition rate and the resonance spacing (dispersion)—generically violates this symmetry, leading to the observed oscillations. 

To understand the impact of anti-parity time (APT) symmetry on the lattice dynamics, we simplify the number of parameters by expressing the Hamiltonian for the odd cavity modes in a dimensionless form. The minimal model requires consideration of only two parameters: The first parameter \(\alpha\), which defines the simplified comb amplitudes – decreasing at a constant rate according to \(A_{\mu}/A_{0}=10^{-\mathcal{R}\cdot\left|\mu\right|/10}, \ \mathcal{R}>0,\ A_{\pm|\mu|}/A_{\pm|\mu+2|} \equiv \alpha\), and reaches $\alpha=0.3548$ in our experiment. The second parameter \(\Delta\tilde{\omega}_\mu\), represents the dimensionless frequency detuning (obtained by dividing by $g_0A^2_0$ where $g_0=6.2[rad/sec]$ in our experiment) of the cavity modes from a rotating frame of reference aligned with the Kerr comb. To the first order of \(\alpha\), we obtained the following form for the lattice Hamiltonian:

\begin{multline}
\
\hat H_{\rm 
lat}=-g_{0}A_{0}^{2}\sum_{\mu\in \rm odd,\mu>0}\left[\frac{1}{2}\left(\Delta\tilde{\omega}_{\mu}+2\right)\hat{c}_{\mu}^{\dagger}\hat{c}_{\mu}+2\alpha\hat{c}_{\mu}^{\dagger}\hat{c}_{\mu-2}\right]\\+\left[\frac{1}{2}\left(\Delta\tilde{\omega}_{-\mu}+2\right)\hat{d}_{\mu}^{\dagger}\hat{d}_{\mu}+2\alpha\hat{d}_{\mu}^{\dagger}\hat{d}_{\mu-2}\right]+\left[\hat{c}_{\mu}^{\dagger}\hat{d}_{\mu}^{\dagger}+2\alpha\hat{c}_{\mu}^{\dagger}\left(\hat{d}_{\mu-2}^{\dagger}+\hat{d}_{2+\mu}^{\dagger}\right)\right]+\text{h.c}+\mathcal{O}\left(\alpha^{2}\right)\\
\label{eq2}
\end{multline}
where $c_\mu\equiv a_{\mu}, d_\mu\equiv a_{-\mu}$ for $\mu>0$, and $d_{-1}\equiv c_1,c_{-1}\equiv d_1$.  We note that the higher-order terms \(\mathcal{O}\left(\alpha^n\right)\) introduce Bragg coupling and pair generation terms to the \(n\)-nearest neighbors, thereby preserving the lattice structure. In this "folded" representation, if the detuning \(\Delta\tilde{\omega}_\mu\) is symmetric around \(\mu=0\), the Hamiltonian in Eq.~\ref{eq2} forms a degenerate pseudo APT-symmetric lattice. This results in an eigenspectrum with eigenvalues that are either purely real or purely imaginary (See Methods~\ref{Methods5} for the symmetry analysis). As was discussed earlier, the real part of the eigenvalues corresponds to parametric amplification, while the imaginary part corresponds to oscillations. Consequently, when APT symmetry exists, extended supermodes which have gain close to threshold may not oscillate at all (purely real eigenvalues). However, in our microcombs, APT symmetry does not exist in general, mainly due to a mismatch  between the repetition rate of the comb and the group-velocity \(\Delta\tilde{\omega}_\mu\). Building on this realization, by adjusting the repetition rate of our microcomb, we can tune the oscillations of the supermode to absolute zero. We confirm the existence of extended non-oscillating supermodes generated by Kerr combs numerically (Methods~\ref{Methods6}), and experimentally as we will now show.

Tuning the repetition rate can be achieved by adjusting the micro-heater voltage \( V \). This adjustment detunes the TE and TM modes differently, thereby altering the location of the mode crossing. Since the phase matching that initiates the 2-FSR microcomb depends on the mode-crossing location, this directly modifies the conditions necessary for the first optical parametric oscillator (OPO) threshold, subsequently affecting the microcomb's repetition rate continuously.

\begin{figure}[H]
\centering
\includegraphics[width=1\textwidth]{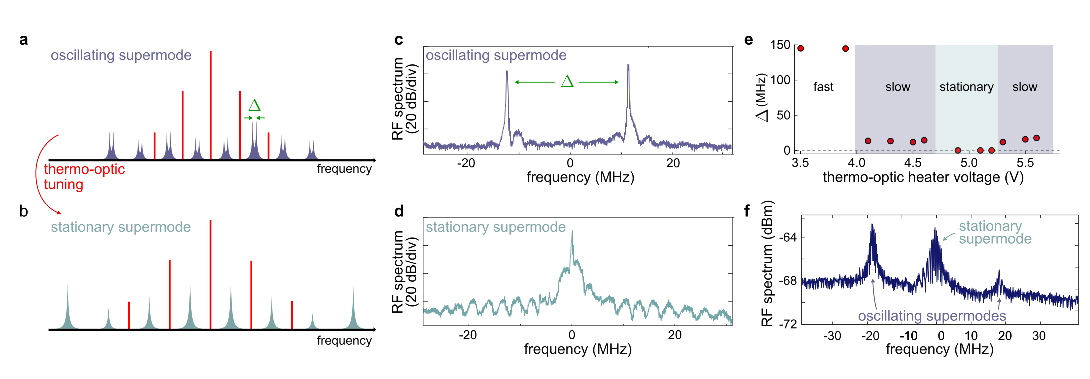}
\caption{{\textbf{Tuning to zero oscillations of the supermodes of vacuum fluctuations} \textbf{a.-b.} Adjusting the voltage of the micro-heater modifies the Kerr microcomb and transitions the supermode's oscillations from non-zero (a) to zero (b). 
\textbf{c.-d.} Experimental tuning of supermode oscillations from non-zero (c) to zero (d) using micro-heaters. The RF spectrum plots are presented above the threshold, showing that the parametric gain peaks precisely at zero, as indicated by the single peak. \textbf{e.} The supermode oscillations as a function of micro-heater voltage demonstrate a discontinuous change from rapid oscillations to slow oscillations, ultimately leading to zero-oscillations. \textbf{f.} RF spectrum below the threshold in a transition region between oscillating and non-oscillating supermodes, illustrating the coexistence of both states with comparable parametric gain.}}\label{fig4}
\end{figure}

To illustrate the transition between oscillating and non-oscillating regimes of the below-threshold light, we continuously tuned the resonator. By varying the voltage applied to the heaters, we observed that repetition rate and the supermode oscillations change, reaching to a non-oscillatory regime as illustrated in Fig.\ref{fig4}(a) and Fig.\ref{fig4}(b). Figure \ref{fig4}(c) presents the RF spectrum of the oscillating supermode at threshold exhibiting a reduction in the oscillation frequency to below 20 MHz, in comparison with the previous range of 100-150 MHz shown in Fig.~\ref{fig3}(c). Further increasing the voltage leads to the complete elimination of the oscillation rate to zero within a finite temperature range, all without the need for fine-tuning (Fig.~\ref{fig4}(d)). Increasing the voltage beyond this point resumes oscillations at approximately 20 MHz, as depicted in Fig.~\ref{fig4}(e), showcasing the different oscillation regimes achieved through tuning.

The three oscillation regimes—fast, slow, and non-oscillating—are predicted through numerical investigations of the lattice dynamics spectrum of supermodes (Methods \ref{Methods6}). Notably, the oscillation's cessation over a finite temperature range is unexpected, suggesting that even when the APT-symmetry is not fully satisfied, but close to being satisfied,  supermodes with real eigenvalues (non-oscillating) remain present. This behavior agrees with classical PT-symmetric system, that exhibit completely real eigenvalues even in non-ideal experimental conditions \cite{regensburger_paritytime_2012}. Our numerical investigations, detailed in Methods \ref{Methods6}, indicate that this robustness is supported by our theoretical model and reveal that non-oscillatory states only begin to oscillate after a significant deviation from APT-symmetric conditions. This finding demonstrates that non-oscillatory multimode squeezed states produced by dissipative Kerr combs can be reliably maintained without stringent control over system parameters.
 
The behavior of non-oscillatory supermodes can be further understood experimentally by observing the transition between a dominant oscillating supermode to a non-oscillating supermode. The transition is not continuous in frequency but includes the two supermodes having similar parametric gain. Figure \ref{fig4}(f) shows exactly that - the below threshold anti-squeezing of two co-existing supermodes at the transition point. At this transition point, small changes to the temperature or input power result in either the oscillating supermode going above threshold, or the non-oscillating going above threshold. If the non-oscillating supermode reaches threshold, additional beat notes would not form as in \ref{fig3}(b) but a single Kerr comb with repetition rate of 1-FSR (See Section 3 in the Supplementary Material for additional experimental measurements of this phenemenon). 

\section{Discussion}\label{sec12}

In this work, we experimentally and theoretically investigated the lattice dynamics of optical multimode vacuum fluctuations generated by spontaneously occurring dissipative Kerr combs. We successfully resolved the RF spectrum of an oscillating supermode that spans 12 resonances. We further explored these oscillations to demonstrate their impact on the secondary threshold process, specifically the generation of classical beat notes after the threshold is exceeded. Notably, we were able to tune the supermode using micro-heaters, effectively extinguishing its oscillation completely within a finite temperature range. We elucidate this phenomenon from the perspective of non-Hermitian lattice dynamics and lattice symmetries.

Examining these fundamental properties, which connect dispersion engineering, emerging phenomena, and multimode vacuum fluctuations, paves the way for observing and engineering all-optical bosonic networks which exhibit multimode quantum squeezing and multimode quantum entanglement. Such phenomena on chip can potentially be generated by injecting or probing various pulses and multimode classical optical drives. Addressing and controlling the dynamics of supermodes is important for utilizing multimode squeezing of Kerr combs in information processing \cite{wang_large-scale_2024,jia_continuous-variable_2025} and sensing \cite{herman_squeezed_2025}. 

Oscillations and other forms of dynamical behavior can naturally arise in applications involving multimode squeezing in cavities, presenting challenges for harnessing squeezing in quantum computing or interferometry below the shot noise limit \cite{gouzien_hidden_2023}. As mentioned in section 4 of the Supplementary information , a double-peak structure can directly reduce the level of squeezing measured with a multi-tone local oscillator. Thus, controlling these dynamics can help mitigate unwanted parasitic effects. Conversely, shaping the RF spectrum has proven a valuable tool for leveraging squeezing in interferometry \cite{mcculler_frequency-dependent_2020}. Consequently, the techniques developed in this work for controlling this spectrum may prove advantageous for similar applications in multimode on-chip setups. 

Moreover, Kerr nonlinearity allows to realize additional exotic states characterized by nontrivial artificial gauge fields, such as topological phenomena \cite{leefmans_cavity_2023,flower_observation_2024}. Other pumping schemes may lead to non-trivial gauge fields in the quadrature space, resulting in quadrature dimers and non-reciprocal quadrature flow \cite{del_pino_non-hermitian_2022,mcdonald_phase-dependent_2018,slim_optomechanical_2024}. Our examination of comb formation and the vacuum noise inherent to Kerr combs provides potential for advancements in comb technology, particularly for those governed by quantum noise, which sets the ultimate limits on comb stability. The concepts articulated here form a bridge that enables exploitation of the rich non-Hermitian lattice physics and the nonlinear effects of Kerr combs to engineer complex quantum resources on chip. The lattice dynamics generated by the Kerr effect, as demonstrated in this work, are also relevant for ultrashort pulses in micro-resonators which also exhibit complex multimode bosonic network dynamics. These phenomena could manifest as below-threshold entangled supermodes, offering new opportunities for controlling quantum light.

\section{Methods}\label{sec11}
\subsection{High-Q 4H-SiC integrated microrings with integrated filters}\label{Methods1}

\begin{figure}[h!]
\centering
\includegraphics[width=1\textwidth]{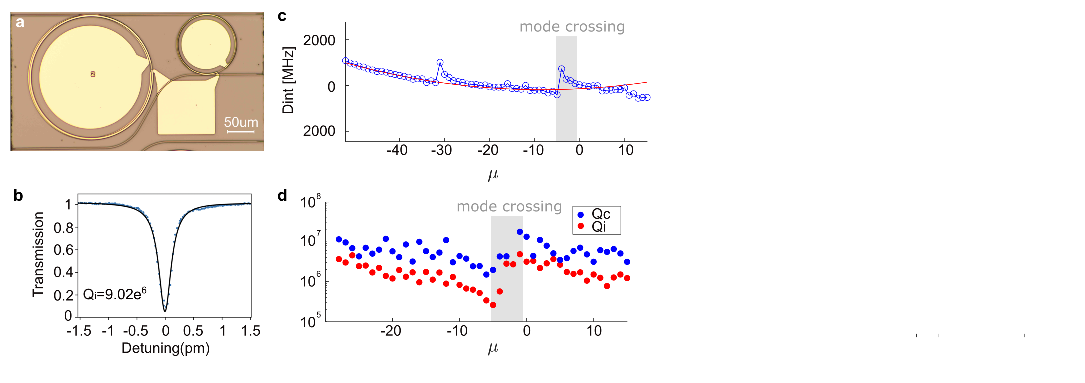}
\caption{\textbf{Micro-ring resonator: fabrication and dispersion. a.} Optical image of the Kerr ring resonator alongside the smaller on-chip filter ring and heaters. \textbf{b.} Normalized transmission over wavelength detuning of a ultra high-Q resonance in our device with intrinsic quality factor of 9 million \textbf{c.} Integrated dispersion of the TE mode of the Kerr ring resonator (blue) and fit to $D_2=1.2~MHz, D_1=152.7$~GHz (red). Dispersion defects are avoided mode crossings with other mode families. \textbf{d.} Cold cavity intrinsic quality factors (Qi - red) and coupling quality factors (Qc - blue) of the resonances in the relevant mode-family. \label{FigM1}
}
\end{figure}

Our devices are etched from a 550~nm thick 4H-SiC-on-Insulator \cite{lukin_4h-silicon-carbide--insulator_2020} and are 1930-nm wide, with diameter of 233-um and capped with SiO\textsubscript{2} (Fig.~\ref{FigM1}(a)). Microheaters are patterned via liftoff of 100~nm platinum with a 5~nm titanium adhesion layer. The filters extinction is smaller than 20dB and they are supplemented with off-chip fiber Bragg filters. Coupling to and from the chip is achieved using lensed fibers, with collection efficiency of up to 50\%. 
The ring resonators exhibit intrinsic quality factors of up to 9 million (Fig.~\ref{FigM1}(b)). The resonators exhibit a $D_2/2\pi$ of 1.2MHz, and an FSR of 153~GHz. Additionally, the ring is multimode, causing mode splitting through modal interactions (Fig.~\ref{FigM1}(c)), which allows a broad range of narrow-bandwidth states to be observed in the devices, such as 2-FSR primary combs, as well as the transition to 1-FSR combs and soliton crystals. The intrinsic quality factors of the mode family used in this work reach 5 million but with average of 1.8 million (Fig.~\ref{FigM1}(d)). The mode splitting is a source  of squeezing reduction due to the mode mismatch at the detector, and the poor output coupling of TM modes. Other factors that reduce the squeezing are the under-coupled to critically coupled resonances and a local oscillator which does not cover 10 cavity modes and mismatch with the oscillating supermode. 
The heaters modify the mode-splitting properties and the index of refraction, thereby altering the integrated dispersion and the pump detuning at threshold which consequently changes the Kerr comb's repetition rate $\Delta\Omega$.

\subsection{Additional details on the 2-FSR microcombs}\label{Methods2}
\begin{figure}[H]
\centering
\includegraphics[width=1\textwidth]{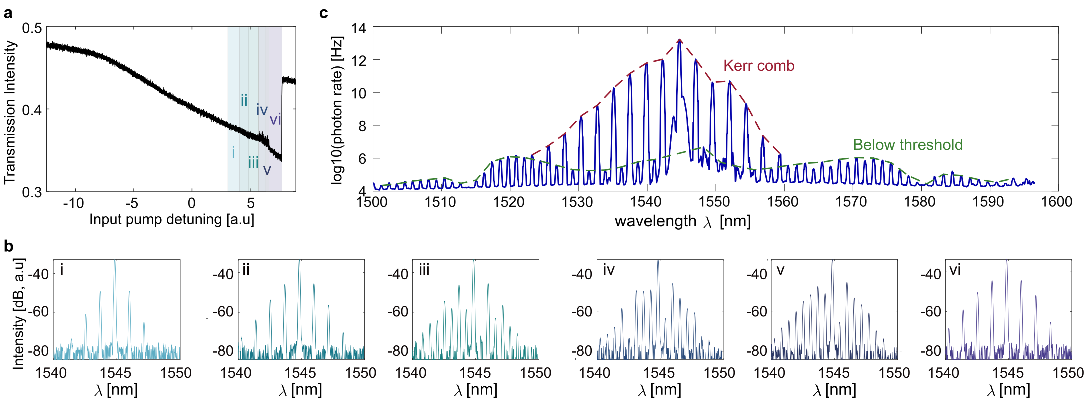}
\caption{\textbf{The experimental optical spectrum of the Kerr microcomb. a.} Transmission as a function of pump detuning. The different colored lowercase Roman numerals correspond to the different stages of the microcomb. \textbf{b.} The spectrum of the different combs specified by lowercase Roman numerals in (a). i and ii correspond to the primary 2-FSR comb which is the main focus of this work, while iii is a 1-FSR comb, and iv-v are chaotic states. Finally vi is a 2-FSR stable soliton crystal state. \textbf{c} Single-photon optical spectrum analyzer (SPOSA) measurement of photon populations in above threshold (red) and below threshold (green) modes of the 2-FSR state. \label{FigM2}
}
\end{figure}

In this section we provide additional details and data on the formation process of our 2-FSR microcombs. Figure \ref{FigM2} presents the experimental evolution of the microcomb. The different stages of the comb evolution are tracked by transmission measurements (Fig.\ref{FigM2}(a)), by an optical spectrum analyzer (OSA) (Fig.~\ref{FigM2}(b)) and by single photon optical spectrum analyzer (SPOSA) (Fig.~\ref{FigM2}(c)). In Fig.~\ref{FigM2}(b) we present the optical spectrum of the comb at different detuning stages. The comb emerges after a first threshold as a 2-FSR primary comb (Fig.~\ref{FigM2}(b)i-ii) and undergoes a non-oscillating threshold (single peak supermode - as in Fig.~\ref{fig4}(d)) to a single 1-FSR phase locked comb (Fig.~\ref{FigM2}(b)iii). Following the threshold, a second oscillating threshold occurs (double peak - see section 3 in the supplementary material) followed by additional thresholds and RF beat notes in the comb (Fig.~\ref{FigM2}(b)iv-v). At this point the RF spectrum broadens and becomes chaotic. Finally, the comb abruptly turns to a 2-FSR soliton crystal (Fig.~\ref{FigM2}(b)vi). The 2-FSR soliton crystal state does not have a long stable 1-FSR soliton state at the end of it. In Fig~\ref{FigM2}(c) the optical spectrum is measured with an SPOSA. This configuration allows to resolve the above and below threshold light in the same measurement sequence due to the high dynamical range. Figure~\ref{FigM2}(c) shows how some of the below threshold light is generated far away from the pump (around 1520 and 1570nm), but also in the center modes (between 1540nm-1550nm), which is where we primarily measure.

\subsection{Additional data on LLE simulation}\label{Methods3}

In order to support the theoretical conclusions and experimental observations, we preformed LLE simulations of a 2-FSR Kerr comb. The LLE simulates the classical comb dynamics \cite{chembo_spatiotemporal_2013}, and allows us to better understand the relation of oscillating and non-oscillating squeezed structures, on the formation of secondary combs.

We start by simulating a dispersion with a mode crossing which means that we locally detune in frequency resonance near the pump. The detuning  yields a comb similar to that of our experiment. The resonance frequency of the mode crossing  changes in response to heating. For the same reason, the detuning of the resonances close to the mode crossing also changes as a function of power in the cavity (i.e it is changing during the evolution of the comb). However, since the effects we describe do not require this behavior, we can observe the main features under constant dispersion. We use experimentally extracted parameters of $D_2/2\pi=1.2MHz$ and extract the dispersion perturbation (from the mode crossing) parameters from the measured dispersion (Methods~\ref{Methods1}). For simplicity, we only perturbe the modes near the pump modes $-2,-1,0,1,2$. For quality factors of the rest of the resonances, we use a constant value of 1.83 million which is the measured average value, and we reduce the intrinsic quality factors at modes distant 17 FSR away from the pump to avoid an opo threshold there as in the measured device. The normalized pump power is $f=5.93$ where $f^{2}=8g_{0}\kappa_{c}P_{in}/\kappa^{3}\hbar\omega_{0}$ and $P_{in}$ is the input power, $\kappa_c$ is the out coupling power loss, $\omega_0$ is the pump angular frequency, and $\hbar$ is the reduced Planck's constant.

Importantly, the formation of our 2-FSR comb is not a direct result of the anomalous dispersion, but it is formed locally due to the mode crossing in the dispersion (see Methods ~\ref{Methods1}). This means that the initial comb formation can be modified by adjusting the  frequency of cavity resonances $-2, 0, 2$. The pump detuning value of the threshold to a 1-FSR comb depends strongly on the dispersion of the odd-numbered modes and the perturbation of modes $-1,1$. By adjusting these parameters, we are able to simulate a comb evolution that is qualitatively similar to what we observe in the experiment. Hence tuning them reduces the oscillations of the supermode until it reaches a region with non-oscillating supermode, similarly to the effect of changing the voltage on the microheaters in the experiment.

\begin{figure}[h!]
\centering
\includegraphics[width=1\textwidth]{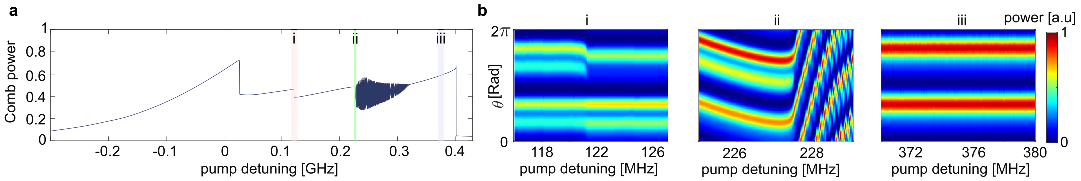}
\caption{\textbf{Intracavity field in real space for different regimes in the comb formation - LLE with perturbed dispersion - non oscillating threshold}. \textbf{a.} Transmission as a function of continuous pump detuning (toward longer wavelengths). This plot is the same as in Fig.~\ref{fig1}(b) but the shaded areas mark different regimes plotted in (b). \textbf{b.} Intracavity field in real space for the transition between the \textbf{(i):} primary comb (2-FSR) and the secondary comb (1-FSR) corresponding to the shaded red region in (a). \textbf{(ii):} the transition from the secondary comb (1-FSR) to a second oscillating 1-FSR comb which oscillates (shaded green region). Experimental evidance of this transition can be found in section 3 of Supplementary Material. \textbf{(iii):} The region of a 2-FSR soliton crystal. \label{FigM3}
}
\end{figure}

Figure \ref{FigM3}(a) shows the comb evolution in the simulation, which is similar to our experiment (Fig.~\ref{FigM2}). The process begins with a 2-FSR primary comb which transitions to a 1-FSR secondary comb. Figure~\ref{FigM3}(b)i demonstrates the transitions from one rolling Turing pattern (2-FSR) to another (1-FSR). Since this transition is non-oscillating, the 1-FSR comb is a single phase locked comb and does not have sub combs (shown experimentally in section 3 in the Supplementary Material). 

The next (third) transition involves the threshold crossing of an oscillating 1-FSR supermode, which produces an RF beat note with the existing 1-FSR comb. The oscillations in this state are evident in both the transmission and the intra-cavity field (Fig.~\ref{FigM3}(b)ii). Finally the system transitions to a 2-FSR soliton crystal (Fig.~\ref{FigM3}(b)iii). To better understand the unique non-oscillating transition from a 2-FSR comb to a single phase locked 1-FSR comb, we explore the comb state at a pump detuning close to this threshold at 0.115~GHz. 
\begin{figure}[h!]
\centering
\includegraphics[width=1\textwidth]{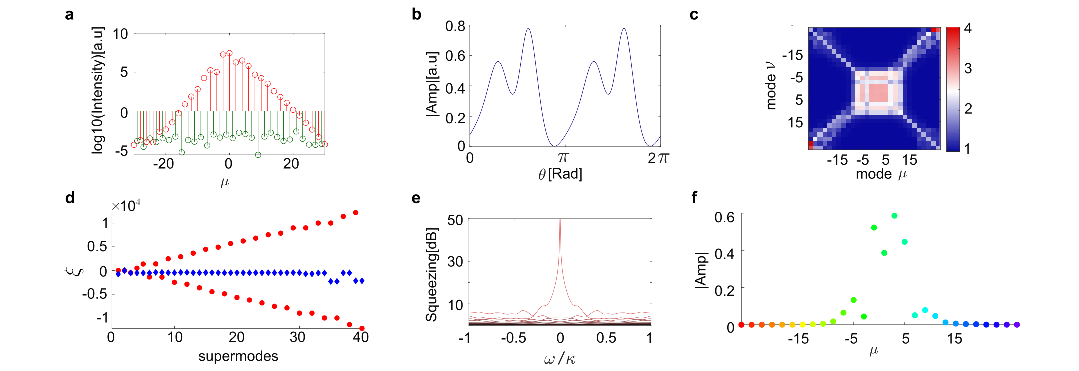}
\caption{\textbf{Examination of the simulated threshold between a 2-FSR comb to a 1-FSR comb at 0.115~GHz detuning}. \textbf{a.} Spectrum of the 2-FSR primary Kerr comb (red) and light in odd-numbered modes (green). \textbf{b.} Intracavity field in real space (rolling Turing) \textbf{c.} $g^{(2)}$ correlations of the below threshold light. \textbf{d.} Eigen-spectrum of the comb showing that a non-oscillating supermode crosses threshold, and that the spectrum also has oscillating supermodes with less parametric gain. \textbf{e.} The squeezing RF spectrum, showing the single peak which correspond with the non-oscillating supermode that crosses threshold in d. \textbf{f.} The spectrum of the most squeezed state in the basis of cavity modes. \label{FigM4}
}
\end{figure}

Figure ~\ref{FigM4}(a-b) shows the comb from Fig.~\ref{FigM3} for the detuning of 0.115~GHz in both the frequency (mode) space and real physical space respectively. Using the obtained comb amplitudes and the hot-cavity dispersion, we calculate the $g^{(2)}$ correlations of each cavity-mode $\mu$ with each cavity-mode $\nu$, which resembles the experimental $g^{(2)}$ in Fig.~\ref{fig2}(c). The oscillating or non-oscillating nature of the supermode is not visible from Fig.~\ref{fig2}(c) and requires the temporal lineshapes. Next, we calculate the quadrature Hamiltonian $\mathcal{M}$ associated with the 2-FSR comb at 0.115GHz. The eigenspectrum is plotted in Fig.~\ref{FigM4}(d), showing that the non-oscillating state (purely real eigenvalue) is approaching threshold. Explicitly, this is evidant from the fact that the supermode with the highest gain (blue dot closest to zero) has zero oscillations (red dot at zero).
Correspondingly, the squeezing structure is a single peak (Fig.~\ref{FigM4}(e)), centered around the pump (Fig.~\ref{FigM4}(f)). When this supermode crosses threshold it does generate RF oscillations.

\subsection{Additional details on the experimental setup}\label{Methods4}
\begin{figure}[h!]
\centering
\includegraphics[width=1\textwidth]{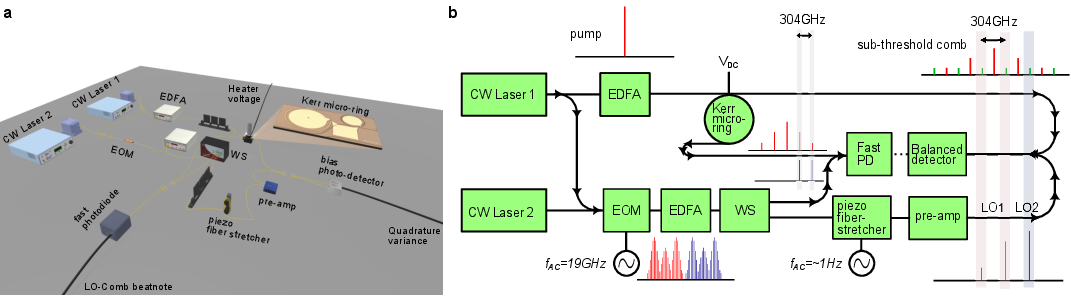}
\caption{\textbf{Principle components of the full experimental setup for quadrature variance measurements}. \textbf{a.} Realistic illustration of the table top setup. \textbf{b.} Schematic diagram of the setup presented in a. the abbreviations are: EDFA- erbium doped amplifier, EOM- electro-optic modulator, WS- waveshaper, PD- photodiode, pre-amp- low-power EDFA,  \label{FigM5}
}
\end{figure}

Here, we provide a more detailed description of the experimental setup. Figure ~\ref{FigM5}(a) illustrates the fiber based experimental setup around the microchip. We utilize two tunable CW telecom lasers named CW laser 1 (Velocity, 1520-1570 nm, TLB-6728) and CW laser 2 (Toptica CTL1500: 1460-1570nm). The lasers are split into two channels. The first channel (upper channel in Fig.~\ref{FigM5}(b)) receives light from laser 1 and after amplification through an EDFA excites the Kerr resonator. The Kerr resonator outputs two channels: the first carries the quantum signal with the filtered Kerr comb directly to the homodyne unit, while the second channel contains the Kerr comb passing through the filter ring to a photodiode.

The second channel (bottom in Fig.~\ref{FigM5}(b)), receives light either from CW laser 1 or CW laser 2 or both. CW laser 1 is used for measuring modes -1 and 1, while CW laser 2 allows measuring other modes and is used for noise spectroscopy in Fig.\ref{fig3}(c). 
To span a 3-FSR range with our LO, we employ an electro-optic modulator (EOSpace LiNbO3 20GHz phase modulator). Subsequently, a programmable WaveShaper (4000A multiport optical processor) transmits specific sidebands that overlap with the modes of the Kerr comb, filtering out all other lines. The frequency tones that overlap with the classical comb are mixed with the teeth that pass through the filter ring in order to track the frequency difference between the Kerr comb and the LO (center of the diagram in Fig.~\ref{FigM5}(b)). The frequency tones of the LO that overlap with the quantum signal are passed through a piezo fiber stretcher that stretches the fiber to change the phase of the LO. This continuous phase ramp is imprinted on the homodyne signal, allowing to distinguish it from other non-phase dependent fluctuations. The signal is then amplified with a low input power EDFA (OptiLab-MSA Pre-Amp EDFA Module, C-band)  and then mixed with the quantum signal from the Kerr comb and fed into a balanced photodetector pair (WL-BPD1GA 1 GHz Dual-Balanced InGaAs Low Noise), to perform the homodyne measurement. 

\subsection{Band model for the lattice dynamics induced by a Kerr comb}\label{Methods5}

As shown in Eq.~\ref{eq2} in the main text, 2-FSR Kerr combs generate 1D lattice dynamics on the frequency axis. The lattice geometry allows reducing the complexity of modeling the physics from a general network of quadratures to a 1D lattice geometry. The goal of this section is to analyze the lattice dynamics by deriving an analytical expression for the band model of the lattice and its Brillouin zone. First, we transform the Hamiltonian in Eq.~\ref{eq2} into the quadrature basis $\vec{R}=(p_{-N},...,p_{N},q_{-N}...,q_{N})^T$ where $p_\mu=\frac{1}{\sqrt{2}}\left(a_\mu^\dagger+a_\mu\right),q_\mu=\frac{i}{\sqrt{2}}\left(a_\mu^\dagger-a_\mu\right)$ and write the quadrature non-Hermitian Hamiltonian $\mathcal{M}$  \cite{gouzien_morphing_2020}, which satisfies the Heisenberg equation: 
\begin{equation}
\
\frac{\partial\vec{R}}{\partial t}=\mathcal{M}\vec{R}\label{eqM1}
\end{equation}

We obtain translation symmetry of the lattice by assuming periodic boundary conditions and flat dispersion  $\Delta\tilde{\omega}_\mu$  in the folded lattice, meaning that $\Delta\tilde{\omega}_\mu=\tilde{\delta}_A\Theta(\omega-\mu)+\tilde{\delta}_B\Theta(\omega+\mu)$ where $\Theta$ is the step function and $\tilde{\delta}_{A(B)}$ are real numbers. We again use the operators $\hat{c}_\mu,\hat{d}_\mu$ and their appropriate quadratures: $\hat{p}_{c,\mu},\hat{q}_{c,\mu},\hat{p}_{d,\mu},\hat{q}_{d,\mu}$.
The difference $\tilde{\delta}_A-\tilde{\delta}_B$ creates an asymmetry in the dispersion $\Delta\tilde{\omega}_\mu$ which removes the anti-parity symmetry, thus it functions as a simplified representation of the mismatch between the comb repetition rate and dispersion $D'_1$. By using the translation symmetry, we transition to the reciprocal space defined by the good quantum number $k$, obtaining a $4\times 4$ Hamiltonian:
\begin{equation}
\
\mathcal{M}_{k}=\left(\begin{array}{cccc}
0 & g_{A,k} & 0 & f_{k}\\
-g_{A,k} & 0 & f_{k} & 0\\
0 & f_{k} & 0 & g_{B,k}\\
f_{k} & 0 & -g_{B,k} & 0
\end{array}\right)\left(\begin{array}{c}
\hat{p}_{d,k}\\
\hat{q}_{d,k}\\
\hat{p}_{c,k}\\
\hat{q}_{c,k}
\end{array}\right)
\label{eqM2}
\end{equation}
where the variables $g_{A(B),k}=\left(\tilde{\delta}_{A(B)}+2\right)+2\alpha e^{ik}+2\alpha e^{-ik}$ originate from the dispersion, Bragg scattering, and self phase modulation, while $f_{k}=-1-2\alpha e^{ik}-2\alpha e^{-ik}$ are pair generation terms, and $\mathcal{M}=\sum_k{\mathcal{M}_k+\mathcal{O}(\alpha^2)}$. The Hamiltonian $\mathcal{M}_k$ has pseudo anti-parity symmetry (implying that the eigenvalues of $\mathcal{M}_k$ are either purely real or purely imaginary) if $\tilde{\delta}_{A}=\tilde{\delta}_{B}$, which shows why there exist an oscillating threshold when $\Delta\tilde{\omega}_\mu$ is not symmetric around $\mu=0$.
\begin{figure}[ht]
\centering
\includegraphics[width=1\textwidth]{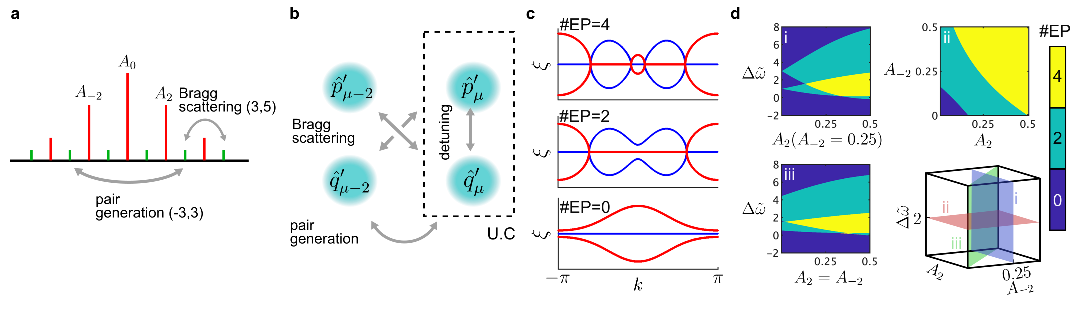}
\caption{\textbf{2-band model of a quadrature lattice induced by a 2-FSR Kerr comb. }\textbf{a}. Illustration of the above threshold comb (red) and below threshold comb (green) in the model, and of the different terms - pair generation and Bragg scattering that couple the different modes. \textbf{b.} The 1D lattice unit cell, in the generalized quadrature basis $\left(\tilde{p}'_\mu,\tilde{q}'_\mu\right)$. \textbf{c.}  Different band topologies (different number of exceptional points) of the Brilloin zone of the lattice in \textbf{b}, top plot corresponds to $A_{\pm2}/A_0=0.25, \Delta\tilde{\omega}=2$, center plot corresponds to  $A_2/A_0=0.25, A_{-2}/A_{0}=0.2, \Delta\tilde{\omega}=2$  and bottom plot to $A_{\pm2}/A_0=0.25, \Delta\tilde{\omega}=0$. \textbf{d.} The number of exceptional points for different 2-FSR Kerr combs on different slices of the 3D parameter space (i-iii). The parameter space is spanned by the two ratios of the sidebands and the overall detuning (bottom right).} \label{FigM6}  
\end{figure}

Therefore, by considering the case in which  $\tilde{\delta}_{A}=\tilde{\delta}_{B}\equiv\tilde{\delta}$ the quadrature Hamiltonian $\mathcal{M}_k$ becomes degenerate. We simplify the treatment further by transitioning to a new quadrature basis in the unit cell according to $\vec{R'}=U\vec{R}$  where $\vec{R'}=\left(p'_{1,k},q'_{1,k},p'_{2,k},q'_{2,k}\right)$ is a generalized quadrature basis, and the unitary is:
\begin{equation}
\
U=\frac{1}{4}\left(\begin{array}{cccc}
1 & 1 & -1 & 1\\
-1 & 1 & -1 & -1\\
1 & 1 & 1 & -1\\
-1 & 1 & 1 & 1
\end{array}\right)
\label{eqM3}
\end{equation}
Applying the change of coordinates with $U$ leads to the following 2 by 2 Hamiltonian:
\begin{equation}
\
\mathcal{M'}_{k}=\left(\begin{array}{cc}
1+2\alpha e^{ik}+2\alpha e^{-ik} & \left(\tilde{\delta}+2\right)+2\alpha e^{ik}+2\alpha e^{-ik}\\
-\left(\tilde{\delta}+2\right)-2\alpha e^{ik}-2\alpha e^{-ik} & -1-2\alpha e^{ik}-2\alpha e^{-ik}
\end{array}\right)
\label{eqM4}
\end{equation}
which is related to $\mathcal{M}_{k}$ of Eq.~\ref{eqM2} by $\mathcal{M}_{k}=U^{\dagger}\mathbf{I}_{2,2}\otimes\mathcal{M}'_{k}U$. The 2-band model of Eq.~\ref{eqM4} represents the most basic lattice quadrature physics and is pseudo Hermitian with APT symmetry. This means that its eigenvalues are either purely real or purely imaginary and it has exceptional points for appropriate lattice parameters. To illustrate the APT symmetry consider the operators $\mathcal{P}=\sigma_x$ and $\mathcal{T}=K$, where $\sigma_x$ is the x Pauli matrix, and $K$ is the conjugation operator. These operators $\mathcal{PT}$ and $\mathcal{M'}_k$ are anti-commuting: $\{\mathcal{PT},\mathcal{M'}_k\}=0$ which shows the APT symmetry.

Due to the degeneracy, we use only two variables: $\left(p'_k,q'_k\right)$, allowing us to explore some of the fundamental physics of our lattices.  Figure~\ref{FigM6} analyzes the 2-band model of Eq.~\ref{eqM4}. Figure~\ref{FigM6}(a) shows different non-linear terms and the structure of the comb (red) and sub-comb(green), while Fig.~\ref{FigM6}(b) presents the folded representation where translation symmetry exists. We stress that in the analysis shown in Fig.~\ref{FigM6} and subsequent simulations we take into acount also next-nearest neighbor terms. These terms preserve the anti-parity time symmetry in the model. Additionally, in the regime of $\alpha<0.5$ the effect of the next nearest neighbor terms on the phase maps were found to be marginal as expected.

Next, we analyze the parameter space of the comb shown in Fig.~\ref{FigM6}(a) where $\alpha<0.5$. In our experiment $\alpha$ reaches 0.355, which is well within this region. We do not address the region of $\alpha>0.5$ or very broad combs in this work, as they require separate study and are beyond the scope of this paper. Regarding the detuning parameter $\tilde{\Delta\omega}$, this parameter varies in the experiment as a function of the pump detuning and the voltage in the microheater. Therefore, we keep it as a free parameter and sweep over it in the widest relevant range.
In this parameter sub-space there can exist 0,2 or 4 exceptional points in the Brillouin zone (Fig.~\ref{FigM6}(c)). We map the 3D parameter space for combs (also for combs with $A_2\neq A_{-2}$) presenting the different band topologies (Fig.~\ref{FigM6}(d)). 

From the phase maps in Fig.~\ref{FigM6}(d), our experimental parameter $\alpha$ (which is also close to the LLE simulation) is expected to have a broad detuning ($\Delta\tilde{\omega}$) region where non-oscillatory states should exist. Additionally, non-oscillatory supermodes are more likely to appear with higher $\alpha$ values. 
We note that having 4 exceptional points is not an inherent maximum;  beyond the presented parameter space, more exceptional points can exist. Additionally, the relative phases between the comb tones can influence the number of exceptional points when more than two pumping amplitudes are considered. Therefore, while we provide the basic mechanisms governing such lattices, they can become significantly more complex. The analysis of the 2-band or 4-band models sheds light on the eigenvector and eigenvalue structures. However, the infinite lattice cannot reach threshold by definition and is therefore  not suitable for studying multi-mode squeezing phenomena on its own, and must be truncated for that purpose. More details on the truncation and its implication on the results of the 2-band model will be discussed in the next section.

\subsection{The finite model analysis}\label{Methods6}

In this section, we continue with the theoretical analysis akin to Methods \ref{Methods5}, focusing on finite lattice dynamics. Similar to structures in physical space, a realistic lattice in the frequency domain is inherently finite. Naturally, dispersion is never completely flat (i.e group velocity dispersion or higher-order terms are never zero), implying that as the lattice extends towards higher and lower frequencies away from the pump, the detuning between the cavity dispersion and the repetition rate increases. This effect reduces the coupling efficiency of Bragg scattering and possibly pair generation, leading to ``soft edges" and resulting in a gradual decrease in transport between frequencies. In contrast, edges can also be made abrupt through engineering of mode crossing or other type of frequency dependent defects \cite{dutt_creating_2022}. These sharp edges can be desirable as they may host confined edge-states in the synthetic frequency dimension \cite{lustig_photonic_2019}. While Kerr resonators can accommodate both options, our focus here is on naturally occurring edges due to non-zero group velocity dispersion. Furthermore, we will numerically demonstrate that the non-oscillating multimode states persist even when dispersion is not completely symmetric and in the finite regime

We start from Eq.~\ref{eq2}, and assume $\Delta\tilde{\omega}_\mu=\omega_p+D'_1\mu+D_2\mu^2$, where $D'_1=\Delta\Omega-D_1$ and $D_1$ represents the FSR. When either $D_2\neq0$ or $D'_1\neq0$ the multimode states become finite in their spectrum, allowing us to analyze them using a finite Hamiltonian $\mathcal{M}$. For this analysis, we assume equal quality factors for all resonances $\Gamma=(\kappa/2) \textbf{I}_{2N\times 2N}$, where, for simplicity, $\kappa$ is the uniform intensity loss rate. Since $\Gamma$ is real and proportional to the unit matrix, it only translates the real part of the eigenspectrum of $\mathcal{M}-\Gamma$, and does not influence the imaginary part of the eigenvalues or the peak separation of the squeezed multimode states. Following the conventions in \cite{gouzien_morphing_2020}, the input-output relation of the system and its Fourier transform are given by:
\begin{equation}
\
\frac{d\mathbf{R}_{out}}{dt}=\left(-\Gamma+\mathcal{M}\right)\mathbf{R}\left(t\right)+\sqrt{2\Gamma}\mathbf{R}_{in}\left(t\right)
\label{eqM5}
\end{equation}
\begin{equation}
\
\textbf{R}_{out}\left(\omega\right)=S\left(\omega\right)\mathbf{R}_{in}\left(\omega\right)
\label{eqM6}
\end{equation}
The linear response matrix $S\left(\omega\right)$ describes in our case how the system amplifies and squeezes the quantum fluctuations and is given by:
\begin{equation}
\
S\left(\omega\right)=\sqrt{2\Gamma}\left(i\omega\mathbf{I}+\Gamma-\mathcal{M}\right)^{-1}\sqrt{2\Gamma}-\mathbf{I}
\label{eqM7}
\end{equation}

To describe the quantum vacuum we solve the frequency dependent linear response.  \cite{gouzien_morphing_2020}. We obtain the Morphing supermodes which describe the dynamical nature of the light. The symplectic form of $\mathcal{M}$ results in a decomposition that is a smooth function of the real parameter $\omega$. Thus, when we apply the Bloch-Massiach decomposition to $\mathcal{M}$ we obtain:

\begin{equation}
\
S\left(\omega\right)=U\left(\omega\right)D\left(\omega\right)V^{\dagger}\left(\omega\right)
\label{eqM8}
\end{equation}
where the matrices $U$ and $V$ are unitary and $D\left(\omega\right)=\text{diag}\left\{ d_{1}\left(\omega\right),...,d_{N}\left(\omega\right)|d_{1}^{-1}\left(\omega\right),...,d_{N}^{-1}\left(\omega\right)\right\}$ corresponds to squeezing and anti-squeezing values of the different quadrature supermodes. The eigenvectors (columns of $U\left\{\omega\right\}$) are morphing with $\omega$ which is detected within the RF spectrum. Thus, the optimization of the local oscillator of the multimode homodyne changes for the same supermode as a function of $\omega$.

Next we would address the OPO thresholds and comb transitions in the context of the complex eigenvalues of the finite model. The eigenvectors of $\mathcal{M}$ are not orthogonal under the usual (Hermitian) inner product which is why the eigenvalues are complex. The complexity of the eigenvalue close but before threshold has profound implications on threshold behavior and secondary comb formation.

First, complex eigenvalues near threshold result in the amplification of oscillating supermodes up to the secondary Kerr comb threshold. The oscillations of the vacuum fluctuations become a double peak in the RF spectrum at each cavity mode leading to multiple RF beat notes after threshold. Thus, if a supermode near threshold has an eigenvalue that is purely real (for example: due to the APT symmetry) RF beat notes will not appear, which will halt the formation of beat notes and consequently chaos until another supermode crosses the OPO threshold. We show this process experimentally in the supplementary section 3. As a side note: APT symmetry does not guarantee that a supermode with purely real eigenvalue will cross threshold, since an entire spectrum (all the supermodes) can also be purely imaginary in the APT case. In this scenario, all supermodes will oscillate and will approach threshold with equal gain until one supermode will break this symmetry and cross threshold. This scenario is simulated and discussed in section 2 of the Supplementary Material. 

Second, measuring the squeezing close to threshold when the modes are oscillating (and non-orthogonal) leads to sub-optimal squeezing measurement, even for optimized local oscillator \cite{gouzien_hidden_2023}.

Next we aim to show more broadly the impact of group velocity dispersion and asymmetry in dispersion relative to the comb on the multi-mode states. We calculate the quadrature eigenspectrum of the sub-threshold light generated by a 2-FSR Kerr comb with fixed $A_\mu$ profile ($\mathcal{R}=2.25$ as previously) and varying $D'_1$. The confinement of the lattice comes from the introduction of non-zero group velocity dispersion $D_2$ and a mismatch between the comb repetition rate and the mismatch $\tilde{D}'_1$. Note that $\tilde{D}'_1$ is the unitless frequency diﬀerence (divided by $g_0A_0^2$) between the cavity FSR and the Kerr comb’s frequency spacing. 

\begin{figure}[ht!]
\centering
\includegraphics[width=1\textwidth]{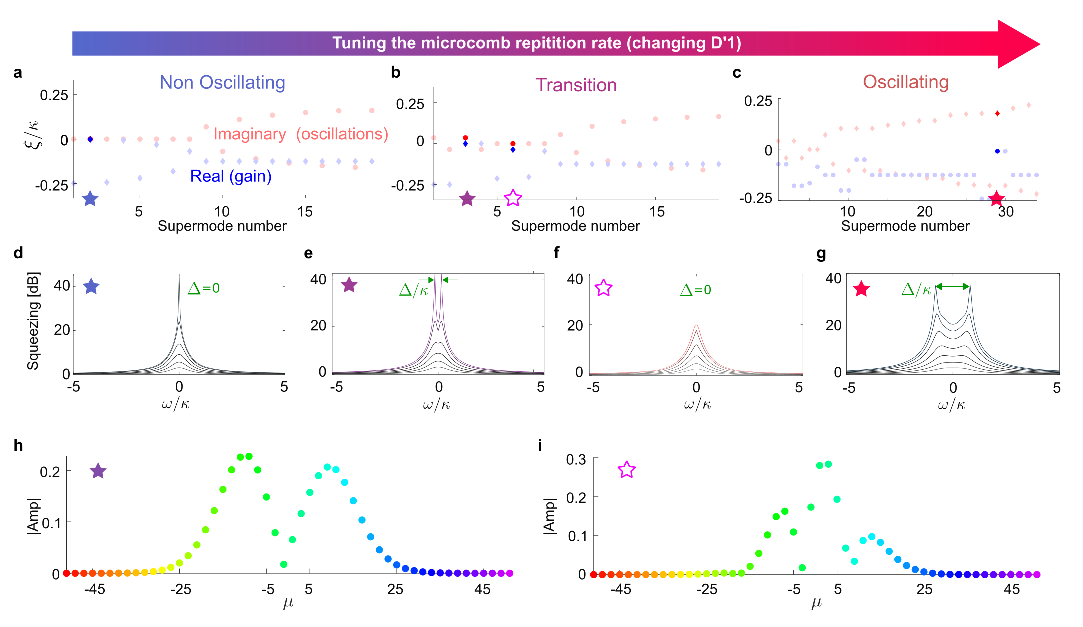}
\caption{ \textbf{Dispersion symmetry and the properties of supermode squeezed states}. \textbf{a.-c.} The real (blue dots) and imaginary (red dots) parts of the eigenvalues of the amplified vacuum supermodes driven by a 2-FSR comb with different values of $D'_1$: 0, 3 and 8~MHz respectively. For larger $D'_1$ the purely real eigenvalues gradually become complex. Highlighted quadrature supermodes are those close to threshold. \textbf{d.-g.} The frequency($\omega$)-dependent squeezing spectrum of the supermodes  close to threshold (highlighted). Each line represents the supermode spectrum for a linearly swept intra cavity power, and the supermodes can be seen to narrow as they approach threshold. The supermode in f. does not narrow asymptotically because e. reaches threshold first. \textbf{h.,i.} The supermode composition as a function of cavity modes of the multimode squeezed states, for the supermodes in e. and f. respectively (colours correspond to the frequencies).}\label{FigM7}
\end{figure}

For $D_2$ and the rest of the parameters we use the value derived from our experiment: $D_2/2\pi=1.2~MHz$. As mentioned $\tilde{D}'_1$ varies as a function of the microheater voltage; therefore, we studied three cases: $\tilde{D}'_1= 0,3,8$ which correspond to the fast, slow, and non-oscillating regimes observed in Fig. 4(e). Additionally, unlike in an infinite lattice, a threshold exists in a finite lattice. To account for this, we include a constant loss value $\kappa$ corresponding to an intrinsic quality factor of 1.74 million. We note that while this value does not alter the qualitative behavior of the dynamics, it does affect it quantitatively. 
Setting $D_2/2\pi=1.2~MHz$, $\mathcal{R}=2.25$, and $Q_i=1.74\cdot10^6$,
Fig.~\ref{FigM7}(a) shows the non-oscillating case in which $D'_1=0$, while Fig.~\ref{FigM7}b(c) are oscillating with $D'_1/2\pi=3(8)MHz$. We adjust the overall comb power in the plots in Fig.~\ref{FigM7} to be close to threshold for detuning $\Delta{\omega}_0=2g\sum_\mu\left|A_\mu\right|^2$. 

To compute the eigenspectrum for a finite model we use Eq.~\ref{eq1} in a finite quadrature basis $(p_1...p_N,q_1...q_N)$ where $p_i$ and $q_i$ are the quadratures of mode $i$, and $N$ is the number of participating cavity modes. We obtain the finite dynamical matrix $\mathcal{M}$ which is non-Hermitian. To study threshold behavior we introduce $\Gamma=\kappa/2 I_{2N\times2N}$, where $\kappa$ is the uniform intensity loss rate.

In Fig.~\ref{FigM7} we analyze the eigenspectrum of $\mathcal{M}-\Gamma$ while varying the parameter $D'_1$, taking into account all orders of $\alpha$. We observe three distinct regimes. For $D'_1=0$ (Fig.~\ref{FigM7}(a)), the eigenvalues of the different supermodes are either purely real or have a real part that equals $-\kappa/2$ resulting from their initial purely real eigenvalue shifted uniformly by $\kappa/2$. Interestingly, due to the multimode nature of the system, even when the symmetry is removed (\textit{i.e.} when $D'_1\neq0$) some states close to threshold remain with imaginary values of 0 (\textit{i.e} the transition region, see Fig.~\ref{FigM7}(b)).  For the highest $D'_1$ in Fig.~\ref{FigM7}(c), the modes close to threshold all have a double peak structure (\textit{i.e.}, imaginary part is non zero).

Each eigenvalue in Fig.~\ref{FigM7}(a-c) represents a multimode squeezed state, residing on the odd-numbered modes. As depicted for example in Fig.~\ref{fig3}(b) under the assumptions taken so far the squeezing spectrum of each supermode uniformly evolves around each cavity mode \cite{guidry_multimode_2023}. The spectral structure around each cavity mode is frequency dependent squeezing, a concept which recently started to be utilized in metrology application through noise spectrum engineering  \cite{PhysRevLett.124.171102}. 
 
To compute the multimode frequency-dependent squeezing structure, we calculate the supermodes of $\mathcal{M}-\Gamma$ for Eq.~\ref{eqM8}, using the Bloch-Messiah decomposition. Specifically, the squeezing RF spectrum of the dominant supermode that reaches the threshold is given by $d_i^{-1}(\omega)$ where $i$ corresponds to the supermode that has the largest real part of the eigenvalue. The RF structure of the supermodes with the largest real part (gain) in Fig.~\ref{FigM7}(a-c) is presented in Fig.~\ref{FigM7}(d-g). 
Supermodes with the highest gain will dominate in total photon number over other supermodes and have the narrowest peaks, being closest to the OPO threshold. The OPO threshold occurs when $Re(\xi)= 0$, where $\xi$ denotes the eigenvalue of the supermode with the highest real part. The frequency spacing in the double peak structure is determined by the imaginary value of the corresponding supermode by $\Delta=2Im{\xi}$. Therefore, supermodes with imaginary value zero, exhibit a single peak structure (``non-oscillating" with respect to the frame of reference of the Kerr comb), while supermodes with imaginary value that is not zero exhibit double-peak spectrum (``oscillating" with respect to the frame of reference of the Kerr comb). 
 
The transition between the two cases shows how the transition between a dominant double peak supermode and single peak supermode is not continuous but is the result of the singe peak supermode increasing in parametric gain compared to the double peak. The transition therefore exhibits both double-peak and single peak evolution for different supermodes with high gain simultaneously. The different plotted lines in each spectral shape in Fig~\ref{FigM7}(d-g) represent the squeezing for different distances from threshold (by changing the overall power of the Kerr comb). Together with an RF frequency-dependent structure, each supermode has a spectral structure in the cavity mode basis, shown for the transition region in Fig.~\ref{FigM7}(h-i). The spectral structure plotted here is the absolute value of the column vector that is extracted from $V\left(\omega=0\right)$. The bandwidth of these supermodes is predominantly determined by the values of $D'_1$ and $D_2$. 

\begin{figure}[ht]
\centering
\includegraphics[width=1\textwidth]{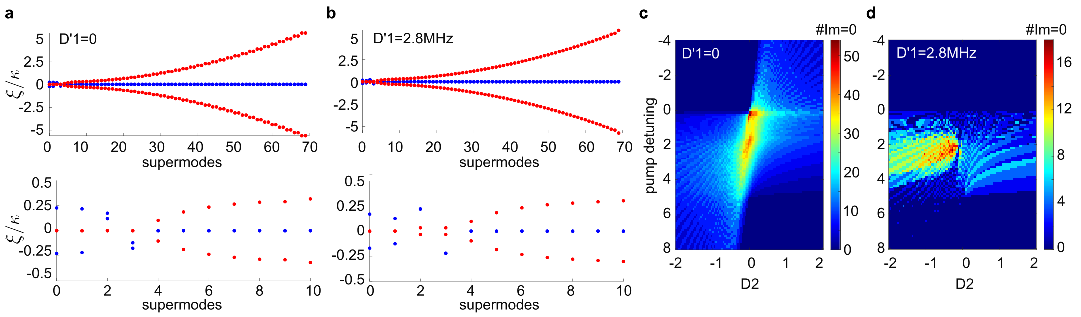}
\caption{\textbf{Existence of non-detuned state in the presence of non-zero $D_2$ and $D'_1$ }\textbf{a}. Top: eigenvalue solution of the quadrature Hamiltonian $\mathcal{M}$ for a 2-FSR Kerr comb with side band drop ratio  of $\mathcal{R}=2.25$, and $D_2/2\pi=1.2~MHz$. Bottom: Zoom-in on the quadrature supermodes with higher-gain showing that all of them are undetuned (imaginary value is 0.) \textbf{b}. Same as a. only with $D'_1/2\pi=2.8~MHz$. \textbf{c.,d.} Numerically counting the un-detuned states as a function of underlying GVD (D2) and detuning of the comb for the symmetric case of $D'_1=0$ and the non-symmetric case $D'_1/2\pi=2.8~MHz$ } \label{FigM8}  
\end{figure}

Next, we wish to discuss the robustness of the non-oscillating supermode. In Fig.~\ref{FigM8}(a) we present a typical complex eigenvalue band structure for the same experimental parameters as before: $D_2/2\pi=1.2~MHz$, $\mathcal{R}=2.25$, and $Q_i=1.74\cdot10^6$. Clearly, dispersion compresses the spectrum bandwidth of the quadrature supermode and the number of participating supermodes. By focusing on the most relevant supermodes, we observe that APT symmetry persists: Supermodes are either purely real (non-oscillating) or purely imaginary (oscillating), with the purely real ones exhibiting higher gain. 

Introducing a mismatch between the cavity FSR and the Kerr-comb frequency spacing eliminates lattice symmetry, yielding more general complex eigenvalue solutions for the $\mathcal{M}$ matrix. When the mismatch is sufficiently small, as depicted in Fig.~\ref{FigM8}(b), we observe that some supermodes remain with purely real eigenvalues while others do not. Figure \ref{FigM8}(c,d) shows the existence of non-oscillating (imaginary part of the eigenvalue is 0) supermodes as a function of $D_2$ and constant pump detuning. Figure \ref{FigM8}(c) enumerates supermodes with purely real eigenvalues for $D'_1=0$, while Fig.~\ref{FigM8}(d) counts the supermodes when $D'_1/2\pi=2.8~MHz$.

\backmatter


\bmhead{Acknowledgements}
We gratefully acknowledge discussions with Edwin Ng, Ryotatsu Yanagimoto, and Kiyoul Yang. This work is funded by the Defense Advanced Research Projects Agency under the QuICC program and by the NSF QuSeC-TAQS Award ID 2326792, as well as the Vannevar Bush Faculty Fellowship from the US Department of Defense and AFOSR Award number: FA9550-23-1-0248. E.L acknowledges the Yad Hanadiv Rothschild fellowship, and the Zuckerman institute Zuckerman fellowship. Part of this work was performed at the Stanford Nanofabrication Facility (SNF) and the Stanford Nano Shared Facilities (SNSF). S. F. acknowledge the support of a MURI project for the U. S. Air Force Office of Scientific Research (Grant No. FA9550-22-1-0339).  We thank NGK Insulators, Ltd. for 4H-SiCOI substrates used to fabricate devices in this work.

\bibliography{sn-bibliography}

\end{document}